\documentclass[12pt,draftclsnofoot,onecolumn]{IEEEtran} 

\newcommand{\subparagraph}{} %
\usepackage{gensymb}

\usepackage[utf8]{inputenc}
\usepackage{titlesec}
\usepackage{xcolor}
\newsavebox{\imgbox}
\usepackage[bottom]{footmisc}
\usepackage[pdftex]{graphicx}
\usepackage[scaled]{helvet}
\usepackage{cite}
\usepackage{epstopdf}
\usepackage{amssymb}
\usepackage{bigints}
\usepackage{xfrac}
\usepackage{amsmath}% for double integral symbol in this template
\usepackage{graphicx}% for figures
\usepackage{multirow}% to allow multiple-row elements in tabular environment
\usepackage[none]{hyphenat}% turn off hyphenation to make text extraction and indexing easier
\usepackage{float}% better control of floating figures and tables
\usepackage[caption=false]{subfig} % Serge: !!!IMPORTANT!!! This keeps the captionsize as it should be!
\usepackage{dblfloatfix}% fix to allow page-wide floats at bottom of page

\usepackage{t1enc}% allows access to various special characters
\usepackage{times}% change font to Nimbus Roman (based on Times Roman)

\begin{document}

%\title{Fully Open-Loop Distributed Beamforming Using Wireless Phase Synchronization}
%\title{Open-Loop Beamforming in Dynamic Distributed Wireless Systems Using High-Accuracy Phase Synchronization}
%\title{Wireless Synchronization for Distributed System Cooperation at the Wavelength Level}
\title{Distributed Radio Frequency Cooperation at the Wavelength Level Using Wireless Phase Synchronization}

\author{Serge R. Mghabghab, Sean M. Ellison, and Jeffrey A. Nanzer% <-this % stops a space

%\thanks{Manuscript received 2020.}
\thanks{This work was supported in part by the National Science Foundation (grant number 1751655), the Defense Advanced Research Projects Agency (grant number N66001-17-1-4045) and the Office of Naval Research (grant number N00014-17-1-2886). The views, opinions, and/or findings contained in this article are those of the author and should not be interpreted as representing the official views or policies, either expressed or implied, of the Defense Advanced Research Projects Agency or the Department of Defense.}
\thanks{The authors are with the Department of Electrical and Computer Engineering, Michigan State University, East Lansing, MI 48824 USA (emails: \{mghabgha, elliso65, nanzer\}@msu.edu).}
}% <-this % stops a space

% The paper headers
%\markboth{IEEE}%
%{Shell \MakeLowercase{\textit{et al.}}: Bare Demo of IEEEtran.cls for Journals}

% make the title area
\maketitle

%\blfootnote{This material is based upon work supported by the Office of Naval Research under #N00014-17-1-2886.}
%\blfootnote{The authors are with the Department of Electrical and Computer Engineering, Michigan State University, East Lansing, MI 48824 USA (email: nanzer@ieee.org).}

%****************************************************************************

\begin{abstract}

Coordinating the operations of separate wireless systems at the wavelength level can lead to significant improvements in wireless capabilities. We address a fundamental challenge in distributed radio frequency system cooperation -- inter-node phase alignment -- which must be accomplished wirelessly, and is particularly challenging when the nodes are in relative motion. We present a solution to this problem that is based on a novel combined high-accuracy ranging and frequency transfer technique. Using this approach, we present the design of the first fully wireless distributed system operating at the wavelength level. We demonstrate the system in the first open-loop coherent distributed beamforming experiment. Internode range estimation to support phase alignment was performed using a two-tone stepped frequency waveform with a single pulse, while a two-tone waveform was used for frequency synchronization, where the oscillator of a secondary node was disciplined to the primary node. In this concept, secondary nodes are equipped with an adjunct self-mixing circuit that is able to extract the reference frequency from the captured synchronization waveform. The approach was implemented on a two-node dynamic system using Ettus~X310 software-defined radios, with coherent beamforming at 1.5~GHz. We demonstrate distributed beamforming with greater than 90\% of the maximum possible coherent gain throughout the displacement of the secondary node over one full cycle of the beamforming frequency.

\end{abstract}

%****************************************************************************

%\begin{IEEEkeywords}
%Antenna arrays, disaggregated antennas, distributed beamforming, frequency synchronization, phased arrays, ranging.
%\end{IEEEkeywords}

\IEEEpeerreviewmaketitle

%****************************************************************************

%\section{Introduction}

Synchronization of distributed electronic systems has long focused on aligning the relative timing between separate nodes. Temporal alignment supports enhanced capabilities in sensor networks and distributed processing, among other applications. As wireless systems become increasingly ubiquitous in society, there has been growing interest in technologies supporting coordination between separate wireless systems such as sensors and communications systems. Of the coordination approaches, those that align the operations of wireless systems at the level of the radio-frequency (RF) wavelength are the most challenging, and also present the most potential for future technology advancement. 
%Coherent distributed antenna arrays, which 
Coordinating separate wireless systems at the wavelength level supports the ability to
disaggregate wireless operations from a platform-centric model to a distributed network of coordinated devices, representing a paradigm shift in wireless system functionality. Without centralized system limitations, greater flexibility can be achieved in terms of scalability, adaptivity to changing conditions or requirements, and lower overall system costs as wireless performance parameters can be directly extended by adding low-cost elements to the network.
%Coherently coordinating separate wireless systems supports distributed beamforming, which will support dramatic improvements in the capabilities of a broad range of wireless applications, from hand-held communications to satellite remote sensing. 
Of the functions enabled by distributed wireless systems, distributed beamforming, where nodes coordinate to steer a phase-coherent wireless signal to a destination, is among the most significant.
Distributed beamforming enables longer link ranges for wireless communications devices by supporting the aggregation of nearby wireless nodes to form a larger array, which can provide significant advantages for users in rural locations and in disaster scenarios. Collections of low-cost UAVs can be used as a distributed array for precision agricultural measurements, and small satellites can be aggregated to perform the operations of a single, larger satellite at a fraction of the cost.
But along with these significant benefits come significant implementation challenges. The principal requirement for distributed beamforming is phase coherence between all nodes in the network, which must be accomplished wirelessly, and is particularly challenging in dynamic distributed systems where the nodes are in relative motion. 
%are systems composed of disaggregated nodes that act coherently to achieve phase-coherent beamforming which leads to improved system performance for applications such as remote sensing, radar, and communications. Coherent distributed arrays are considered as more efficient and robust in comparison to single-platform system, but this advantage comes at the cost of complex synchronization requirements. 

Node synchronization has been approached using closed-loop architectures where the distributed nodes coordinate using feedback from the targeted location or from other external systems\cite{tu2002coherent, brown2008time}. Such approaches simplify the node coordination process and make coherent beamforming possible as long as the network is operating in a cooperative environment, where feedback from external sources is possible, but limits beamforming to locations where feedback is available, preventing beamforming to arbitrary directions. For applications such as radar and remote sensing, beamforming to locations without coherent feedback is necessary, and thus closed-loop architectures are not feasible. Even for some communications applications, feedback may not be available; for example, in scenarios where individual radios have insufficient power to establish a link to a base station, a set of radios may need to form a beam without base station feedback. To support arbitrary phase-coherent wireless operations from distributed antennas systems, open-loop architectures, where the network self-aligns the electrical states of the nodes without external feedback, are required \cite{nanzer2017open, 8058723}. This approach requires more complex coordination but alleviates the restrictions imposed on the applications of the system, making it possible to use a distributed network for any wireless operation, including remote sensing, radar, and communications. 

Coordinating open-loop distributed wireless networks requires the nodes to synchronize in phase, frequency, and time to support and maintain beamforming.
%Achieving coherent open-loop distributed arrays requires the synchronization of time, phase, and frequency. 
Phase coherence guarantees a coherent summation of the received signals; in open-loop networks, such phase coherence necessitates high-accuracy ranging techniques to estimate the delays needed to correct for the relative phase shifts between the distributed nodes \cite{8678474, tu2002coherent, 7317806, 7050381}. Furthermore, maintaining phase coherence is not possible unless all the nodes are frequency locked; since no physical connection exists between the nodes, the signals generated on each node are derived from independent oscillators. Over relatively short time frames, the relative frequencies of the oscillators drift, requiring continuous wireless frequency synchronization to achieve coherent operation \cite{8889331, syllaios2007reconfigurability, 6949699}. Finally, while supporting and maintaining phase alignment, the most challenging coordination task in open-loop distributed wireless networks, signals with information must also be time aligned to ensure sufficient overlap of the pulses or symbols at the destination. The required timing accuracy is thus dependent on the information rate, and is typically on the order of nanoseconds, which can support a greater absolute error than phase alignment which is typically on the order of picoseconds \cite{6624252, 7421334}. 

%\begin{figure}[t!]
	%\centering
	%\includegraphics[width=0.49\textwidth]{Figures/CDA11}
	%\caption[Optional caption]{\textbf{Distributed beamforming in wireless systems.} An open-loop coherent distributed array wirelessly coordinated using a primary/secondary approach. Secondary nodes perform frequency synchronization and phase adjustment in reference to the primary node. Once the secondary nodes localize in reference to the primary node, it is possible to adjust their relative phases to achieve coherent beamforming.}
	%\label{fig:OV-1}
%\end{figure}

In this paper, we demonstrate the first open-loop distributed beamforming at 1.5~GHz using fully wireless coordination in a dynamic distributed two-node system in relative motion. High-accuracy inter-node ranging for phase alignment is achieved using a spectrally sparse, scalable ranging waveform while frequency synchronization is supported by transmitting a two-tone modulated frequency reference which is demodulated using a self-mixing receiver. We demonstrate a two-node distributed system based on software-defined radios (SDRs), transmitting 1.5 GHz continuous wave (CW) signals towards end-fire, which will later be shown to be the most challenging steering angle. Experimental results demonstrate the ability to maintain greater than 90\% of the ideal beamforming gain in the presence of relative node motion.

\textbf{}

\textbf{Phase Coordination For Distributed Beamforming}

\begin{figure}[t!]
	\centering
	\textbf{a}
	\includegraphics[width=0.49\textwidth]{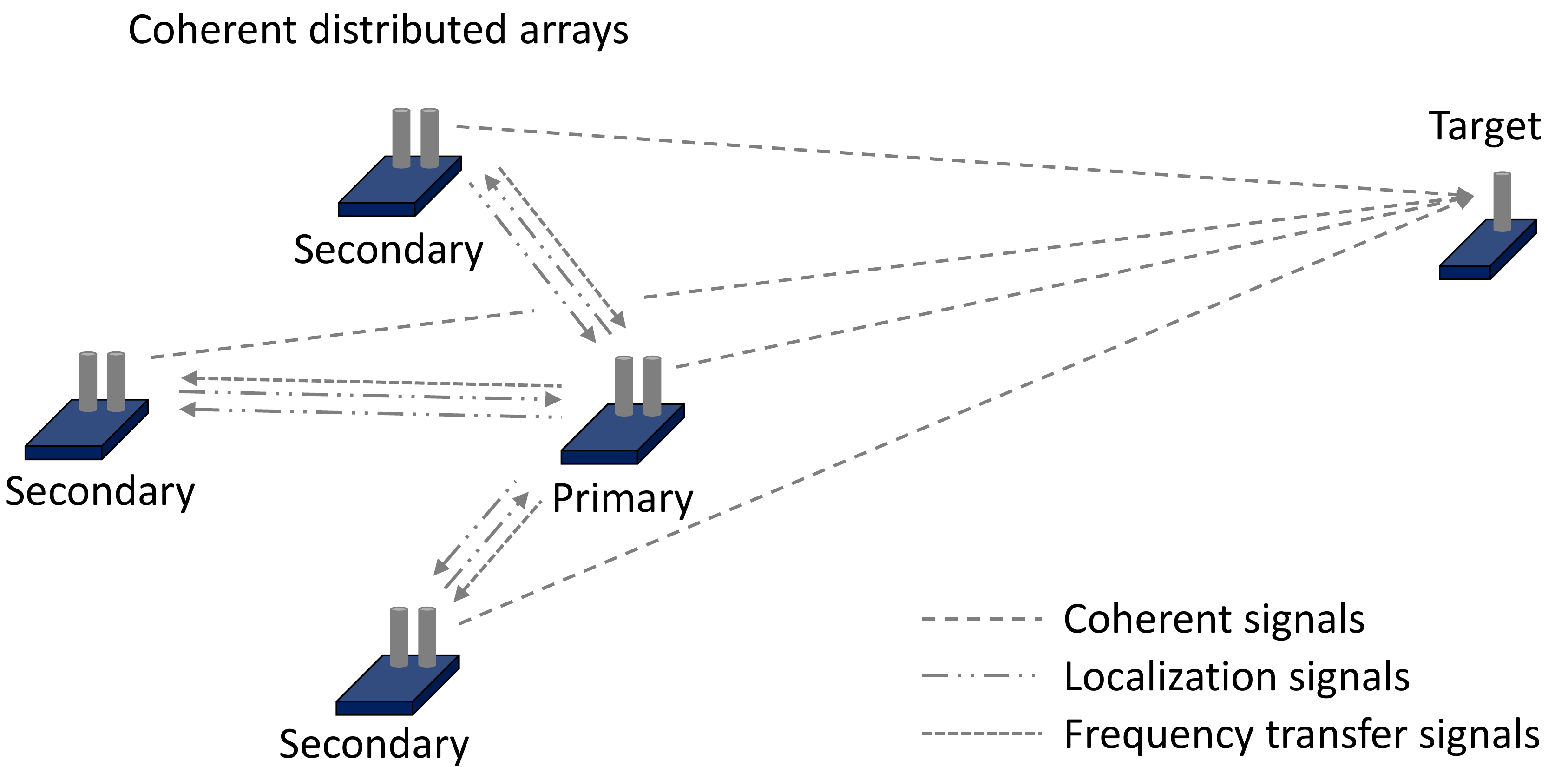}
	
\textbf{}
	\textbf{b}
	\includegraphics[width=0.75\textwidth]{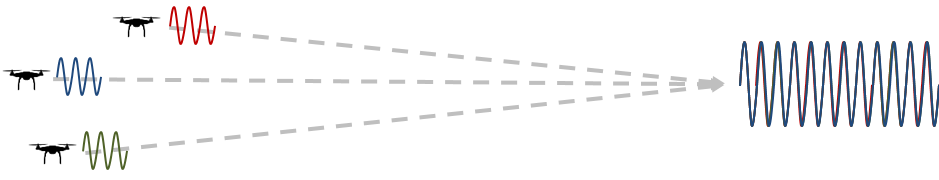}
		
\textbf{}
	\textbf{c}
	\includegraphics[width=0.3\textwidth]{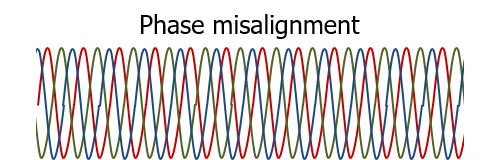}	
	\includegraphics[width=0.25\textwidth]{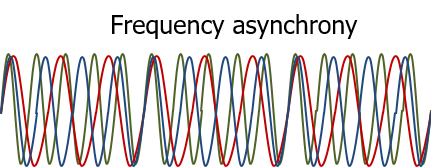}
	\includegraphics[width=0.3\textwidth]{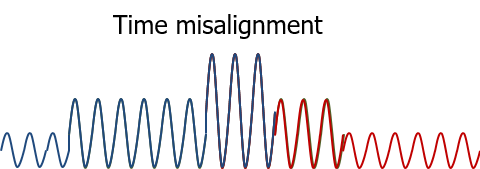}
	\caption[Optional caption]{\textbf{Distributed beamforming in wireless systems.} \textbf{a.} An open-loop coherent distributed wireless system coordinated using a primary/secondary approach. Secondary nodes perform frequency synchronization and phase adjustment in reference to the primary node. 
	\textbf{b.} Once the secondary nodes localize in reference to the primary node, it is possible to adjust their relative phases to achieve coherent beamforming. \textbf{c.} The fundamental coordination challenges are alignment of the phase, frequency, and time of the transmitted waveforms. Time alignment is dependent on the inverse of the information bandwidth, which is generally orders of magnitude less stringent than the accuracy necessary for phase and frequency alignment. In this work we demonstrate a framework for obtaining and maintaining phase and frequency synchronization, and demonstrate fully open-loop distributed beamforming.}
	\label{fig:CDA}
\end{figure}

Of the coordination aspects necessary in distributed systems, described in Fig. \ref{fig:CDA}, frequency and phase synchronization are the most fundamental, as without both, coherent transmission is impossible. 
%We are interested in ensuring coherent beamforming of systems that require phase alignment and frequency synchronization such as the example shown in . 
The requirements for time alignment are dependent on the bandwidth of the transmitted waveform, whereas more stringent requirements are needed for the phase alignment since it must be synchronized at the level of the wavelength of the carrier frequency \cite{8378649}. In \cite{nanzer2017open} it was shown that to achieve 90\% of the maximum possible coherent gain with a high probability, it is sufficient to ensure that the ranging accuracy is less than or equal to $\lambda_c/15$ when the steering angle is considered as a random variable, where $\lambda_c$ is the wavelength of the transmitted carrier. The coherent gain is equal to the power obtained from the beamformed signals relative to the maximum achievable coherent power. A modified metric is used in this paper to assess the ranging requirements to take into account the effects of the frequency synchronization circuit. In general, the desired ranging accuracy will be on the order of centimeters or millimeters for microwave and millimeter-wave operation. Many waveforms were characterized in \cite{weiss1983fundamental, weinstein1984fundamental, nanzer2016bandpass} to determine the best achievable ranging accuracy, and it was shown that a higher accuracy can be obtained when the spectral power is concentrated around the edges of the achievable bandwidth. The ideal case of a split-band signal is represented by a two-tone waveform where each signal is an impulse function in the frequency domain. The main challenge of using two-tone waveforms for ranging is that the resulting measurement, while optimally accurate, is highly ambiguous, making it challenging to differentiate multiple point target returns. Thus, the two-tone waveform tends to be used mostly in cooperative systems, such as between nodes in distributed wireless systems, where measurements can be disambiguated more easily. A two-tone stepped frequency waveform (TTSFW) was introduced in \cite{9057428}, where it was shown that such a waveform could take advantage of the high accuracy two-tone waveform and use it in a scalable, unambiguous manner by time duplexing pulse sequences with modified starting points.

Apart from range estimation, wireless frequency synchronization is crucial for phase alignment, and has been studied for communications applications for a number of years.
%Wireless frequency synchronization was studied for multiple applications such as in communications and wireless sensing, among many others. 
In Orthogonal Frequency Division Multiple Access (OFDMA), frequency synchronization ensures low levels of inter-channel interference and enables signal orthogonality. Frequency synchronization is typically accomplished in a two-way process: time alignment is first ensured and then the frequency is locked using specific data packets \cite{8649676, morelli2007synchronization}. In wireless sensing, closed-loop architectures were implemented for frequency synchronization \cite{seo2008feedback}. Open-loop architectures were also implemented such as in \cite{wang2014carrier}, but these approaches are slow and require significant processing. Coupled-oscillators \cite{ponton2017stability} and optically-locked voltage controlled oscillators \cite{yang2014picosecond} were implemented for wireless frequency synchronization, however the performance of such systems highly depends on the separation distance, and in the case that obstacles are present in between the transmitter and receiver, frequency synchronization might not be possible. Decentralized approaches were used for open-loop wireless frequency synchronization \cite{9028079}, nevertheless, tracking the phase shift produced by the nodes displacements is not easily done.

%****************************************************************************
\textbf{}

\textbf{Inter-Node Range Estimation}

Monitoring the positional change produced by the relative motion of the nodes within a coherent distributed system is essential to enable coherent beamforming, since any change in relative phase of the transmitted signals will impact the coherent summation of the signals. In a primary/secondary hierarchical architecture, every secondary node must track its position in comparison to a reference point, which is the location of the primary node. % Therefore, all the secondary nodes need to continuously estimate their relative position to the primary node using an accurate radar.% A repeater can be installed on the primary node, so the received ranging pulses can be amplified and retransmitted to the secondary nodes; this will increase the ranging accuracy since the radar returns will have a higher power leading to a better SNR and ranging accuracy.
 Based on the radar returns, the relative phase shifts can be calculated and then accounted for. The radar waveform that is employed needs to support accurate delay estimation, must be able to unambiguously determine range, and it must be scalable so it can be used by multiple nodes simultaneously to reduce update latency.

%\subsection{ranging Waveform}
The TTSFW ranging waveform used in this work is based on a spectrally sparse two-tone waveform, which has been proven to have optimal performance as a time delay estimator, in which the obtainable accuracy is dependent on the spectral separation of the tones \cite{7801084}. The TTSFW uses a two-tone format on a pulse-by-pulse basis where the frequencies are monotonically increased with time and is expressed as \cite{9057428}
\begin{equation}\label{eq:1}
s_t(t)=\frac{1}{N}\sum_{n=0}^{N-1}\text{rect}\left(\frac{t-nT_r}{T}\right)\left(e^{j2\pi f_1t}+e^{j2\pi f_2t}\right)e^{j2\pi n \delta f t}
\end{equation}
where $f_1$ is the lower tone of the two-tones per pulse, $f_2$ is the upper tone, $\delta f$ is the frequency step, $N$ is the number of pulses, $T_r$ is the pulse time duration, and $T$ is the nonzero portion of the pulse duty cycle. The frequency step is given by $\delta f=\frac{BW}{2N-1}$ where $BW$ is the total waveform bandwidth. The higher of the two frequencies is given by $f_2=f_1+\Delta f$ where $\Delta f = N \delta f$. 

\begin{figure} [t!]
	\centering
	\textbf{a}
	\includegraphics[width=0.5\linewidth]{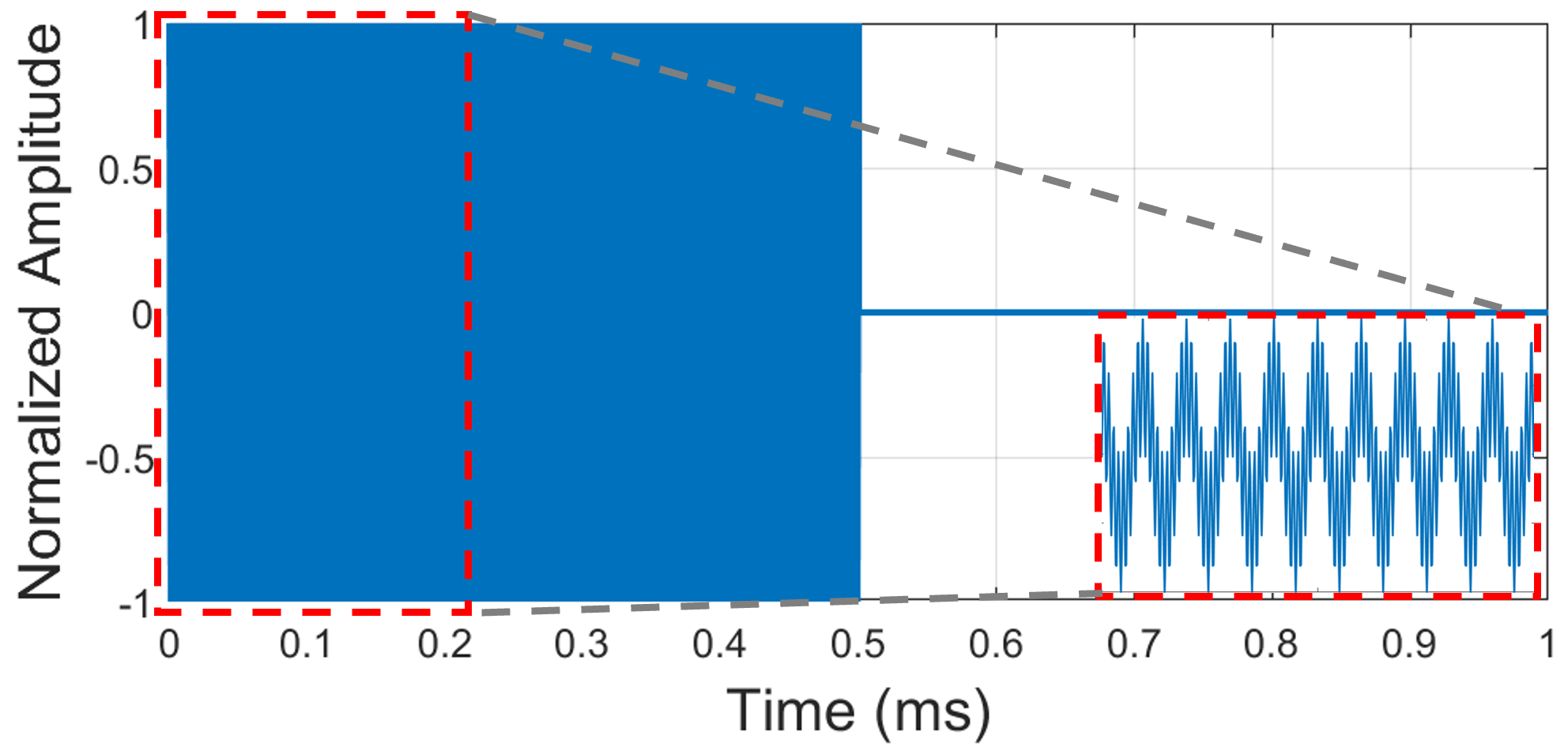}
	
	\textbf{}
	
	\textbf{b}
	\includegraphics[width=0.5\linewidth]{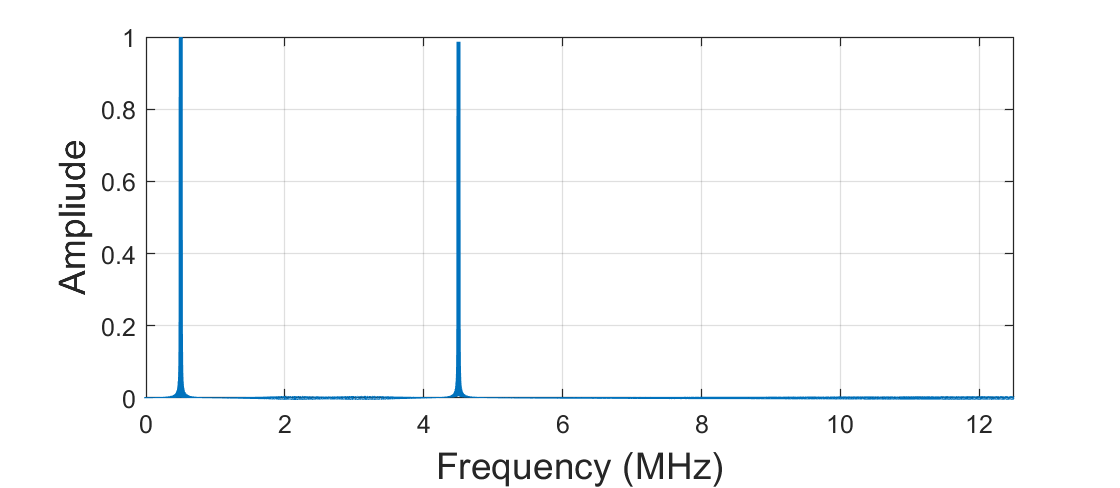}
	
	\textbf{}
	
	\textbf{c}
	\includegraphics[width=0.49\textwidth]{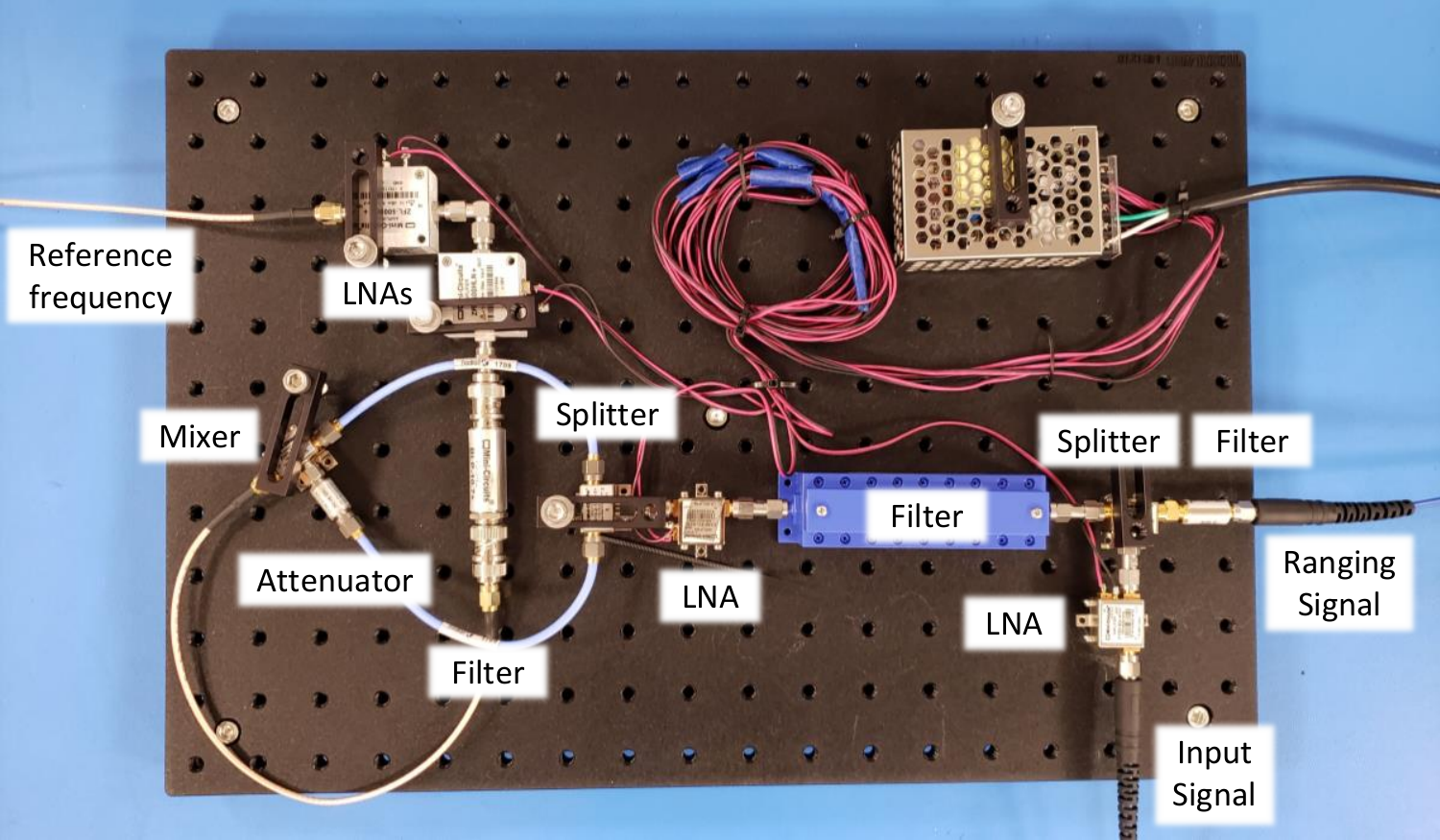}
	%\caption[Optional caption]{\textbf{Experimental implementation of the frequency locking circuit.} Once amplified, the input to the presented circuit is split to provide signals for simultaneous ranging and frequency synchronization.}
	
	\caption{\footnotesize{\textbf{Ranging waveform and frequency synchronization circuit.} \textbf{a.} Time domain representation of the ranging waveform. \textbf{b.} Frequency content of the ranging waveform. \textbf{c.} Experimental implementation of the frequency locking circuit.} Once amplified, the input to the presented circuit is split to provide signals for simultaneous ranging and frequency synchronization.}
	\label{fig:waveform}
\end{figure}

%%%%%%%%%%%%%
%\subsection{Estimation Ability} \label{sec:EstimationAbility}

The estimation ability of a function transmitted from one node, retransmitted back from a second, and received by the first node is given by the Cramer-Rao lower bound (CRLB) \cite{7801084} defined by the Fourier parameters of the domain of interest. Therefore, the ability of a function to estimate the time delay between the two nodes is given by
\begin{equation}\label{eq:tau}
\sigma_\tau^2\geq\frac{1}{\frac{2 E}{N_0}(\zeta_f^2-\mu_f^2)}
\end{equation}
where $\sigma_\tau^2$ is the variance of the time estimate, $E$ is the signal energy, $N_0$ is the noise power per unit bandwidth, and $\mu_f$ and $\zeta_f^2$ are the first and second moments of the frequency spectrum respectively. $E/N_0 = T \cdot BW_n \cdot \text{SNR}$, where $T$ is the non-zero time duration of the ranging pulse, $BW_n$ is the noise bandwidth, and SNR is the signal to noise power ration. The spectrum of the TTSFW is fundamentally symmetric about the mean so the estimation ability becomes a function of only the second moment.
The range estimation accuracy of the TTSW can be derived from \eqref{eq:tau} yielding
\begin{equation}\label{eq:2}
\sigma_x^2\geq\frac{c^2}{8\frac {E}{N_0}\zeta_f^2}
\end{equation}
where $\sigma_x^2$ represents the variance of the position estimate, $c$ is the speed of light, and the factor of four comes from the two way propagation that is seen by typical radar measurements. For the TTSFW, the second moment of the spectrum is given by \cite{9057428}
\begin{equation}\label{eq:3}
\zeta_f^2=\pi^2\left(\frac{BW}{2-\frac{1}{N}}\right)^2+\frac{(2\pi BW)^2}{N(4N^2+4N+1)}\sum_{n=0}^{N-1}n^2
\end{equation}
which for a 4 MHz bandwidth and the aforementioned waveform parameters can be derived as $\zeta_f^2\Big|_{N=1}=\pi^2BW^2$ $=~1.5791\times10^{14}~\text{Hz}^2$.
We then evaluate \eqref{eq:tau} and \eqref{eq:2} at an SNR of 30~dB, matching the SNR observed in wireless measurements described later. This level of SNR is feasible by using a cooperative ranging system which experiences only one-way propagation losses when compared to traditional radar systems that experience two-way propagation losses. This is made possible by the primary node repeating back the ranging information to the secondary nodes with added transmission gain.

%$N$ is chosen based on the number of the secondary nodes, where $N$ pulses can be used to service $N!$ connections.

The number of pulses in the TTSFW is dependent on the number of unique ranging connections made, where $N$ pulses can be used to service $N!$ connections. A system consisting of two nodes, one primary and one secondary, has only one unique connection and therefore only a single pulse is need. An image showing the time domain and frequency domain representation of the waveform used in this paper can be seen in Fig. \ref{fig:waveform}. The waveform was generated using LabVIEW on a computer that was connected to an Ettus X310 SDR using a 10 Gb network interface. The sampling frequency was set to 25 MHz, while the frequencies of TTSFW were selected as $f_1 = 500$ kHz and $f_2 = 4.5$ MHz and the pulse repetition interval had a duration of 1 ms with a 50\% duty cycle. It should be noted that with a sampling rate of 25 MHz a larger $f_2$ could be chosen which would improve the ranging performance. However, it was found that with upper tones close to the Nyquist frequency, distortion in the discrete matched filter output was too coarse to accurately detect small changes in position. By reducing the value of $f_2$, the estimation ability is reduced but results in a more finely sampled representation of the matched filter and therefore improved the ability to measure small positional changes.

The processing gain from using a matched filter process is equivalent to the time-bandwidth product $N\cdot T \cdot BW_n$, where $N$ is the number of pulses which is one, $T$ is equal to 0.5~ms in this work, and $BW_n$ is the noise bandwidth which, since no filtering was preformed, was 12.5~MHz. The processing gain is equivalent to 38 dB giving a total post processing SNR of 68 dB. From \eqref{eq:tau} the timing variability can be solved for as $\sigma_\tau^2=5.066\times10^{-22}$ $s^2$ and from \eqref{eq:2} the positional accuracy can be found as $\sigma_x=3.4$~mm. 

%****************************************************************************
\textbf{}

\textbf{Wireless Frequency Synchronization}

Frequency synchronization is accomplished using an adjunct self-mixing analog circuit \cite{8889331, SergeIMSBeamform}. This architecture utilizes a self-mixing circuit to demodulate a reference frequency from a continuous two-tone signal. 
% Two-tone signals are used in unambiguous high-accuracy ranging, thus one waveform can be used to perform both ranging and frequency synchronization. In case the desired ranging accuracy requires a higher bandwidth, it was shown in \cite{8870258} that three-tone signals can be used for simultaneous ranging and frequency synchronization with almost no effect on the ranging accuracy. In this proposed architecture, the primary node transmits the synchronization signal while the secondary node(s) are responsible for the range estimation and phase adjustment, which made it sufficient to transmit a two-tone frequency synchronization signal.
In this work we are interested in disciplining the oscillators of secondary node(s) using a 10 MHz reference frequency, which correlates to a frequency separation of 10 MHz of the two tones.
% This reference signal is demodulated from a two-tone signal with 10~MHz tone separation.
Once demodulated, the reference 10 MHz is fed to the phase-locked loops (PLLs) internal to the X310 SDRs.
%As mentioned, the secondary nodes were responsible for reference signal demodulation and ranging, the circuit used for this process is shown in Fig. \ref{fig:FLC}.
To reduce spectral interference between frequency synchronization and ranging operations, signals were each transmitted at different frequency bands. The circuit is shown in Fig. \ref{fig:waveform}, present on the secondary nodes, is utilized to amplify the received signals and split them to two separate processes, one to perform ranging and the other for frequency synchronization. An extra low-noise amplifier (LNA) was required for the frequency synchronization signal so adequate power was delivered to the mixer. Once amplified and filtered, the frequency synchronization signal was split into two signals and self-mixed, resulting in a demodulated reference signal whose frequency is equal to the two-tone separation. A 10~dB attenuator was placed at the input of the RF port of the mixer to ensure linear operation. The output was filtered to reject any %undesired frequency component such as the harmonics of the 10~MHz.
spurious frequencies created by the mixer. Extra LNAs were used at the end of the circuit to drive the power to be between 0 and 15 dB which is required for the PLLs of the X310s.

%\begin{figure}[t!]
	%\centering
	%\includegraphics[width=0.49\textwidth]{Figures/FLC}
	%\caption[Optional caption]{\textbf{Experimental implementation of the frequency locking circuit.} Once amplified, the input to the presented circuit is split to provide signals for simultaneous ranging and frequency synchronization.}
	%\label{fig:FLC}
%\end{figure}

The normalized two-tone frequency synchronization signal is represented by
\begin{equation}\label{eq:ttSync}
s_{sync}(t) = e^{j2\pi f_{r1} t} + e^{j2\pi f_{r2} t}
\end{equation}
where $f_{r1}$ and $f_{r2}$ are the two tones separated by the reference frequency 10~MHz. The normalized LO and RF inputs to the mixer are
\begin{equation}\label{eq:VLO}
V_{RF}(t) = \sin\left(2 \pi f_{r1} t + \phi_1\right) + \sin\left(2 \pi f_{r2} t + \phi_2\right)
\end{equation}
\begin{equation}\label{eq:VRF}
V_{LO}(t) = \sin\left(2 \pi f_{r1} t + \phi_3\right) + \sin\left(2 \pi f_{r2} t + \phi_4\right)
\end{equation}
where the phases $\phi_1$, $\phi_2$, $\phi_3$, and $\phi_4$ are obtained from the distance separating the primary and secondary nodes, with $\phi_3 = \phi_1 + c_1$ and $\phi_4 = \phi_2 + c_2$, where $c_1$ and $c_2$ are obtained from the mismatch in cable lengths connecting the splitter to both the RF and LO inputs. In the case where there is a minimal or no mismatch in cables length, $c_1$ and $c_2$ are negligible, maximizing the output power of the resulting signal.

The mixer output $V_{IF}$ is composed of the fundamental frequencies $f_{r1}$ and $f_{r2}$ %, their summation, subtraction, and the resulting harmonics
 along with their spurious frequencies \cite{maas1986microwave}. To suppress all the unwanted frequencies, a filter was added. Since in our case the 10~MHz signal was the lowest output frequency, a low-pass filter with a 10.7~MHz cutoff frequency was used. After filtering, the output signal is represented by
\begin{equation}\label{eq:VIF}
V_{IF}(t) = \cos\left(2\pi\left(f_{r2}-f_{r1}\right)t+\phi_5\right)
\end{equation}
where $\phi_5 = \phi_2-\phi_1$.

%At a later point, tracking the changes of phase constant $\phi_5$ is going to be one of the main goals; since the phase shift produced at the output reference frequency signal is going to dictate the phase shift $\Delta \phi_{ref}$ at the oscillator's output.
The phase shift $\phi_5$ is mainly caused by the inter-node separation $\Delta d_{IN}$% between the two-tone transmitter and the receiver of the frequency locking circuit
. The change in the phase constant of the output, $\Delta \phi_{ref}$, can be tracked by estimating the phase shifts for the two received frequencies $\Delta \phi_1$ and $\Delta \phi_2$,
\begin{equation}\label{eq:Dp1}
\Delta \phi_1 = - \frac{f_{r1} \cdot \Delta d_{IN} \cdot 360^{\circ}}{c}
\end{equation}
\begin{equation}\label{eq:Dp2}
\Delta \phi_2 = - \frac{f_{r2} \cdot \Delta d_{IN} \cdot 360^{\circ}}{c}
\end{equation}
\begin{equation}\label{eq:Dpref}
\Delta \phi_{ref} = \Delta \phi_2 - \Delta \phi_1 = - \frac{f_{ref} \cdot \Delta d_{IN} \cdot 360^{\circ}}{c}
\end{equation}
where $f_{ref} = f_{r2} - f_{r1}$ is the frequency of the reference signal. $\Delta \phi_{ref}$ is negative when the separation between the two nodes increases because the received signal is delayed, thus the output phase reflects this delay.

Once the shift of the output phase is determined, the phase changes of the transmitted signal can be tracked. Signals generated from the SDR's internal clock retain clocking phase information and therefore, the effect of the phase changes of the oscillators on the transmitted carrier can be represented by
\begin{equation}\label{eq:Dpc1}
\Delta \phi_{c_1} = \frac{f_c}{f_{ref}} \Delta \phi_{ref} 
\end{equation}
where $f_c$ is the carrier frequency.

It is important to note that \eqref{eq:Dpref} is only dependent on the tone separation $f_{ref}$ and not the actual frequencies $f_{r1}$ and $f_{r2}$. This shows that the phase shift related to the displacement distance will eventually only depend on $f_{ref}$ in \eqref{eq:Dpref}. As a result, the final phase shift reflected on the transmitted carrier will only depend on the transmitted carrier $f_c$ as shown in \eqref{eq:Dpc1}. This finding will affect the required ranging accuracy for coherent beamforming which will be discussed later.

%****************************************************************************
\textbf{}

\textbf{Distributed Beamforming With Wireless Phase Synchronization} \label{sec:summation}

\begin{figure}[t!]
	\centering
	\includegraphics[width=0.49\textwidth]{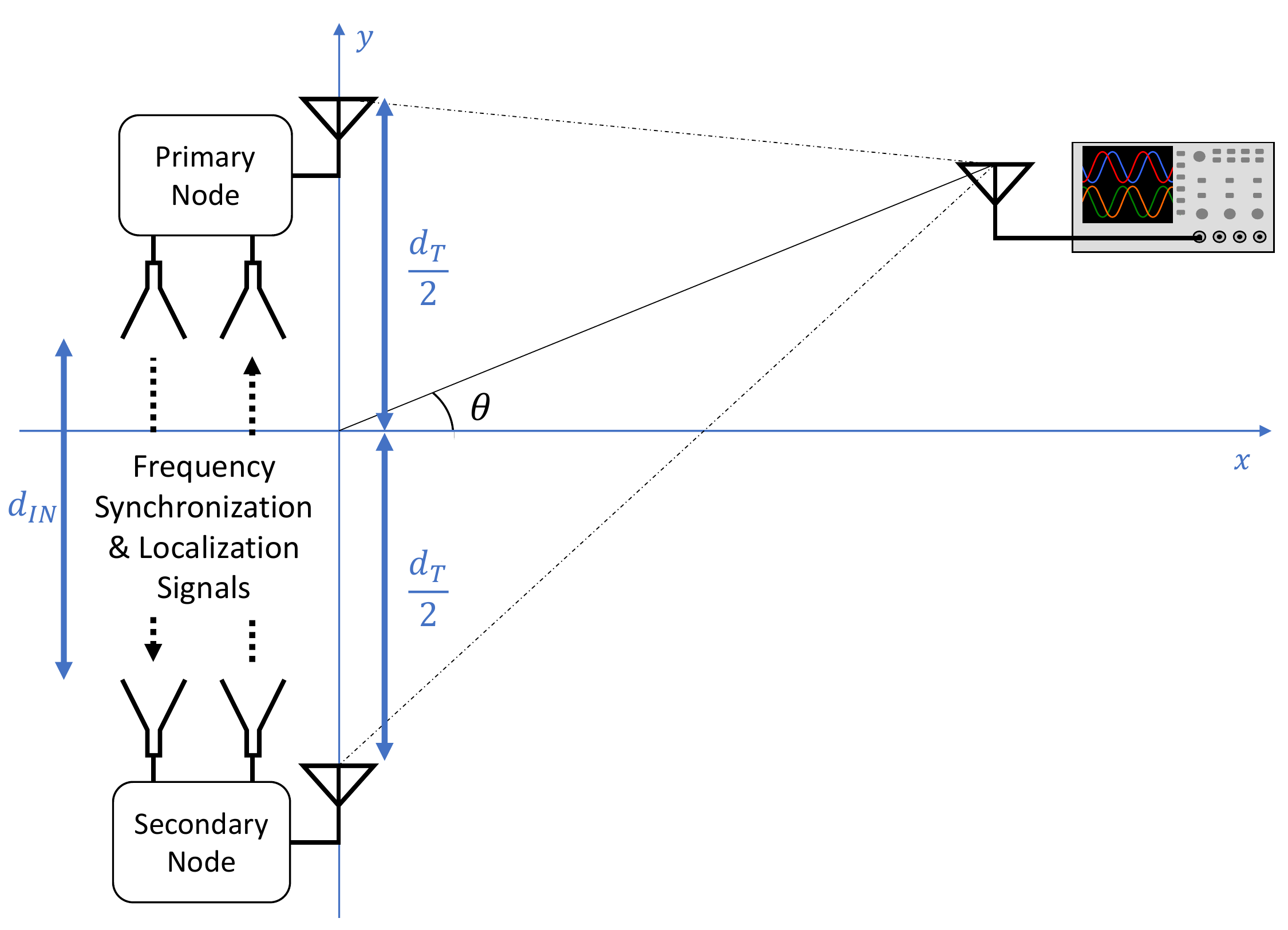}
	\caption[Optional caption]{\textbf{Schematic of the distributed wireless system.} The two-node system consisted of a primary and secondary node, beamforming at an arbitrary angle to a target receiver represented here by a signal analyzer.}
	\label{fig:Nodes_schematic}
\end{figure}

The phase shift $\Delta \phi_{c_1}$ is generated by the displacement of the primary antenna transmitting the two-tone synchronization waveform to the receiver of the frequency locking circuit. Another phase shift, $\Delta \phi_{c_2}$, can now be examined for the secondary node which is proportional to the displacement $d_T$ of the antennas performing the beamforming. Fig. \ref{fig:Nodes_schematic} shows an example of a two-node distributed system beamforming to an arbitrary angle $\theta$, where the target is assumed to be in the far field. In a similar fashion to traditional phased arrays, the transmission from the secondary node will require a phase shift so that the signals constructively cohere at the target destination. By calibrating the beamforming phases before operation at a specific separation $d_T$, the phase shift observed from the secondary node will be the relative change in displacement $\Delta d_T$. The phase shift $\Delta \phi_{c2}$ can then be determined by 
%\begin{equation}\label{eq:Dpc2}
$\Delta \phi_{c_2} = - \frac{1}{\lambda_c}\Delta d_T \sin(\theta) \cdot 360^{\circ}$
%\end{equation}
where $\theta$ is the beam steering angle and $\lambda_c$ is the wavelength of the beamforming frequency.
The total phase shift observed from moving the beamforming transmitters as well as the frequency synchronization antennas is expressed by $
%\begin{equation}\label{eq:Dpc}
\Delta \phi_{c} = \Delta \phi_{c_1} + \Delta \phi_{c_2}$.
%\end{equation}
Coherent beamforming can then be achieved, once the system is calibrated, by using the ranging data to track the motion of the distributed nodes and the resultant $\Delta \phi_{c}$ can be accounted for.

In a similar fashion to \cite{nanzer2017open} the summation of the beamforming signals can be expressed as
\begin{equation}\label{eq:sr1}
s_{R}(t) = \sum_{n=1}^{N} h_n A_n (t-\tau_n) e^{j \left[ 2 \pi f_c(n)  t + \Delta\phi_c(n) + \phi_0(n)\right]}
\end{equation}
where $h_n$ is the complex valued response for the $n^{\mathrm{th}}$ propagation channel. The amplitude of the $n^{\mathrm{th}}$ signal is represented by $A_n$, $\tau_n$ is the time delay between the $n^{\mathrm{th}}$ transmitted signals, and $\phi_0(n)$ the initial phase constant of the received signals which is calibrated manually in a one-time calibration process. To have a coherent summation of the received signals, the values of $f_c(n)$ need to be equal. This is achieved through frequency synchronization. To ensure that the value of $\tau_n$ is small or negligible, there needs to be an adequate overlap of transmitted information which is achieved through time alignment. In this work, CW signals are transmitted at the frequency $f_c$, thereby enabling beamforming without requiring time alignment. The resulting coherent signal can then be represented by
% Finally, the constant phase shifts $\phi_0(n)$ are calibrated out manually. The calibrated version of the coherent signals is expressed as
%\begin{equation}\label{eq:sr2}
%s_{calibrated}(t) = \sum_{n=1}^{N} h_n A_n (t) e^{j \left[ 2 \pi f_c  t + \Delta\phi_c(n) + \delta\phi(n)\right]}
%\end{equation}
%A final step for ensuring coherent summation would be to account for all the phase shifts caused by the displacement of the secondary nodes in relation to the primary node. The coherent summation is expressed as 
\begin{equation}\label{eq:sr}
s_r(t) = \sum_{n=1}^{N} h_n A_n (t) e^{j \left[ 2 \pi f_c t + \Delta \phi_c(n) - \Delta\phi_{c{_e(n)}} + \delta\phi(n) \right]}
\end{equation}
where $\Delta\phi_{c{_e(n)}}$ represents the ranging error in the estimated phase shift caused by the motion of the nodes and $\delta\phi(n)$ contains the instantaneous phase and frequency errors that are a result of the imperfections in the frequency locking circuit and the generated synchronization signals, as well as interference.

%****************************************************************************
%\section{Ranging Accuracy Requirements for a Desired Coherence Level} \label{sec:accuracy}

\begin{figure}[t!]
	\centering
	\includegraphics[width=0.49\textwidth]{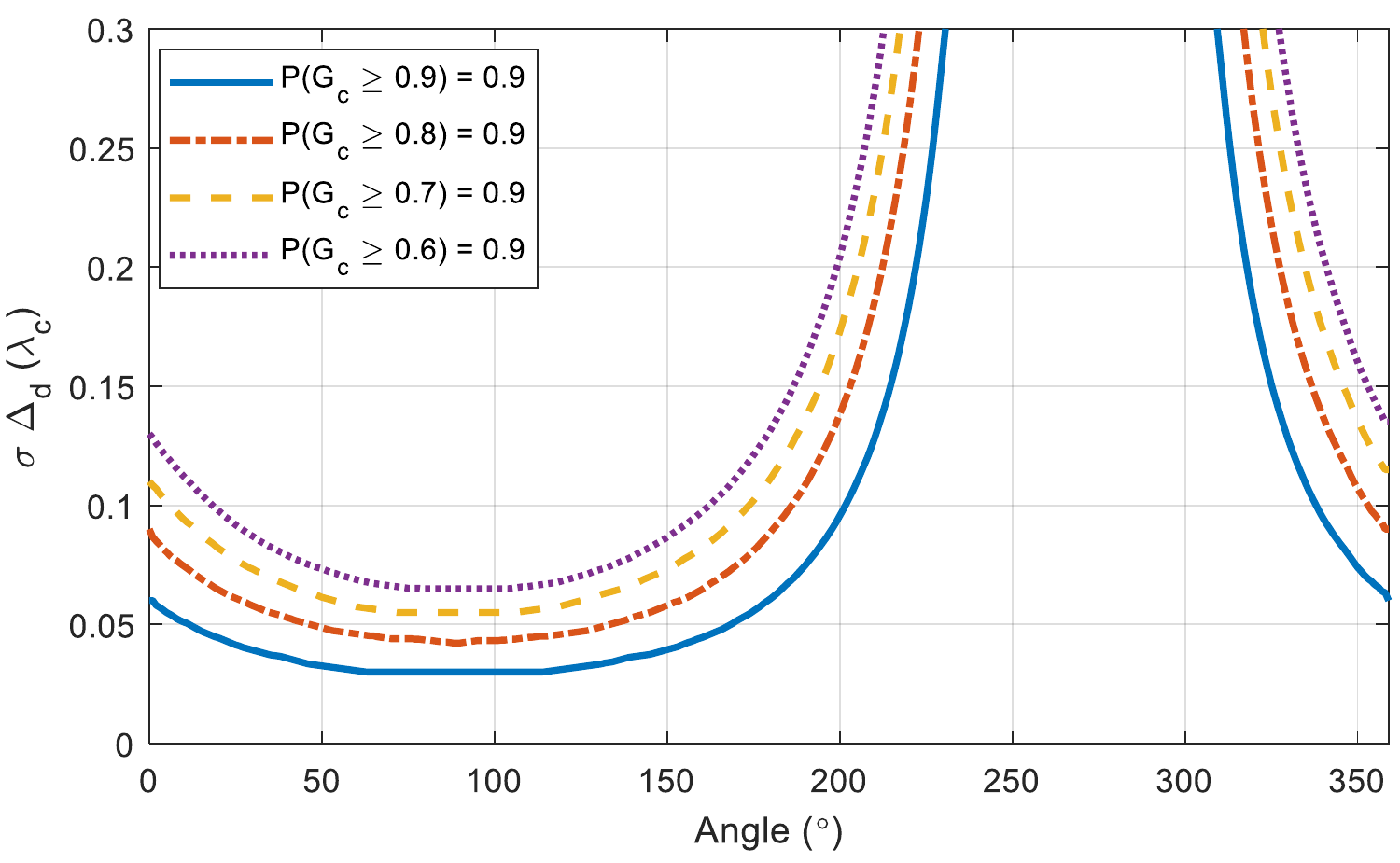}
	\caption[Optional caption]{\textbf{Phase coordination requirements.} 50,000 Monte Carlo simulation showing various threshold values for achieving coherent gain. The probabilities were estimated as a function of steering angle and ranging standard deviation. The strictest requirements for ranging uncertainty appear at $\theta=90^{\circ}$. At $\theta=270^{\circ}$, the contributions from $\Delta \phi_{c_1}$ and $\Delta \phi_{c_2}$ negate each other, making any ranging error tolerable for this beam steering angle.}
	\label{fig:PGC}
\end{figure}

%Evaluation of the dependence of coherent gain to system errors is given by the proportionality factor $G_c$
To evaluate the distributed beamforming performance we use as a metric the coherent gain $G_c$, which is expressed as
%\begin{equation}\label{eq:Gc}
$G_c = \frac{\left|s_r s_r^*\right|}{\left|s_i s_i^*\right|}$
%\end{equation}
where $0 \leq G_c \leq  1$, $s_r$ is the coherent signal with errors present,
 %given in \eqref{eq:sr}, 
and $s_i$ is the ideal summation of the received signals which is expressed as
%\begin{equation}\label{eq:si}
${s_i(t) = \sum_{n=1}^{N} h_n A_n (t) e^{j 2 \pi f_c t}}$.
%\end{equation}
To evaluate the required ranging accuracy, $G_c$ is evaluated parametrically as a function of the ranging standard deviation. The probability to achieve coherent gain above a certain threshold $P \left(G_c \geq X\right)$ is evaluated, where $0 \leq X \leq  1$ is a fraction of the ideal coherent gain. 50,000 Monte Carlo iterations were generated where the distances $\Delta d_{IN}$ and $\Delta d_{T}$ are equal -- which is the case portrayed in Fig. \ref{fig:Nodes_schematic} -- and set to a random number. The values of $h_n$, $A_n$, and $\delta\phi(n)$ were set to 1, 1, and 0 respectively. % At a later stage,  $\delta\phi(n)$ can be added as a compounded error as shown in \cite{nanzer2017open}.
 Fig. \ref{fig:PGC} shows the simulated results where the standard deviation of the range estimates,  $\sigma_{\Delta_d}$, are expressed as a function of wavelength of the coherent signal. The threshold values $X= $ 0.6, 0.7, 0.8, and 0.9 were investigated for a probability $P \left(G_c \geq X\right) = 90\%$.
The results show that beamsteering to end-fire ($\theta = 90^{\circ}$) requires the most stringent ranging accuracies while at $180^{\circ}$ the requirements become more relaxed. This is the case because at $90^{\circ}$, $\Delta \phi_{c_1}$ and $\Delta \phi_{c_2}$ add constructively, while at $180^{\circ}$, $\Delta \phi_{c_1}$ and $\Delta \phi_{c_2}$ add destructively. The minimum standard deviation requirement for the ranging measurements is for $P \left(G_c \geq 0.9\right) = 90\%$ at $90^{\circ}$ which is equal to 0.03 $\lambda_c$; the achievable coherent frequencies using these requirements are given by
%\begin{equation}\label{eq:highestFc}
$f_c \leq \frac{0.03~c}{\sigma_x}$.
%\end{equation}
For a 1.5~GHz signal,  $P \left(G_c \geq 0.9\right) = 90\%$ can be achieved for a maximum ranging uncertainty of 6 mm. This condition ensures that there will be at most 0.5 dB degradation from the ideal coherent gain assuming that the only uncertainty is related to the ranging measurements. We note that the ranging system described earlier supports ranging accuracies below this requirement.

%****************************************************************************
\textbf{}

\textbf{}

\textbf{}

\textbf{Distributed Beamforming Experiment}

%\begin{figure*}[t!]
	%\centering
	%\includegraphics[width=0.99\textwidth]{Figures/Setup_BlockDiagram2}
	%\caption[Optional caption]{\textbf{Schematic of the two-node open-loop distributed beamforming experiment.}  %For the wired signals summation, the antennas M2 and S2 were replaced by cables and were directly connected to the oscilloscope.
	%}
	%\label{fig:SetupBD}
%\end{figure*}

%\begin{figure}[t!]
	%\centering
	%\includegraphics[width=0.49\textwidth]{Figures/expSetup}
	%\caption[Optional caption]{\textbf{Experimental setup of the open-loop coherent distributed array.} Set in a semi-enclosed antenna range, this image shows the target, primary, and secondary nodes. }
	%\label{fig:expSetup}
%\end{figure}

\begin{figure}[t!]
	\centering
	\textbf{a}
	
	\includegraphics[width=0.99\textwidth]{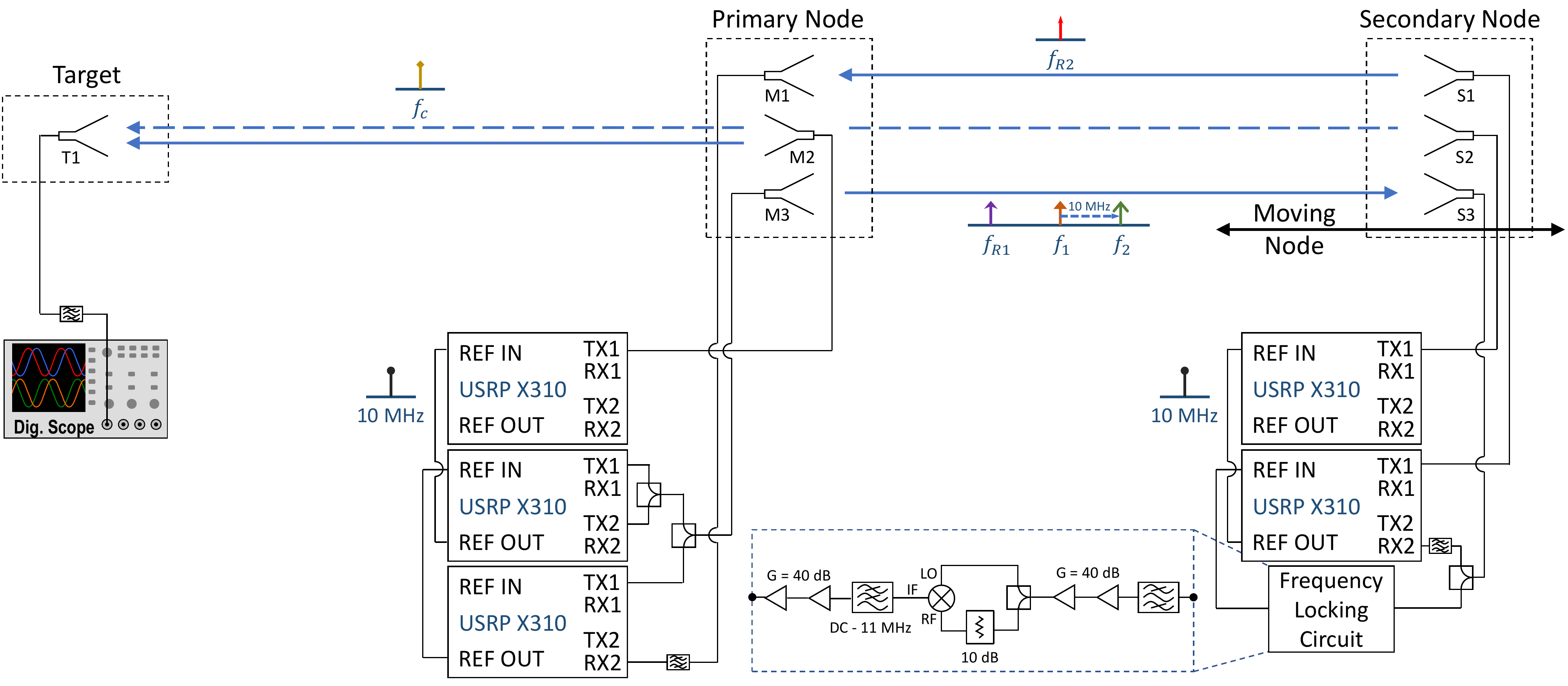}
	%\caption[Optional caption]{\textbf{Schematic of the two-node open-loop distributed beamforming experiment.}
	
	\textbf{}
	
	\textbf{b}
	\includegraphics[width=0.35\textwidth]{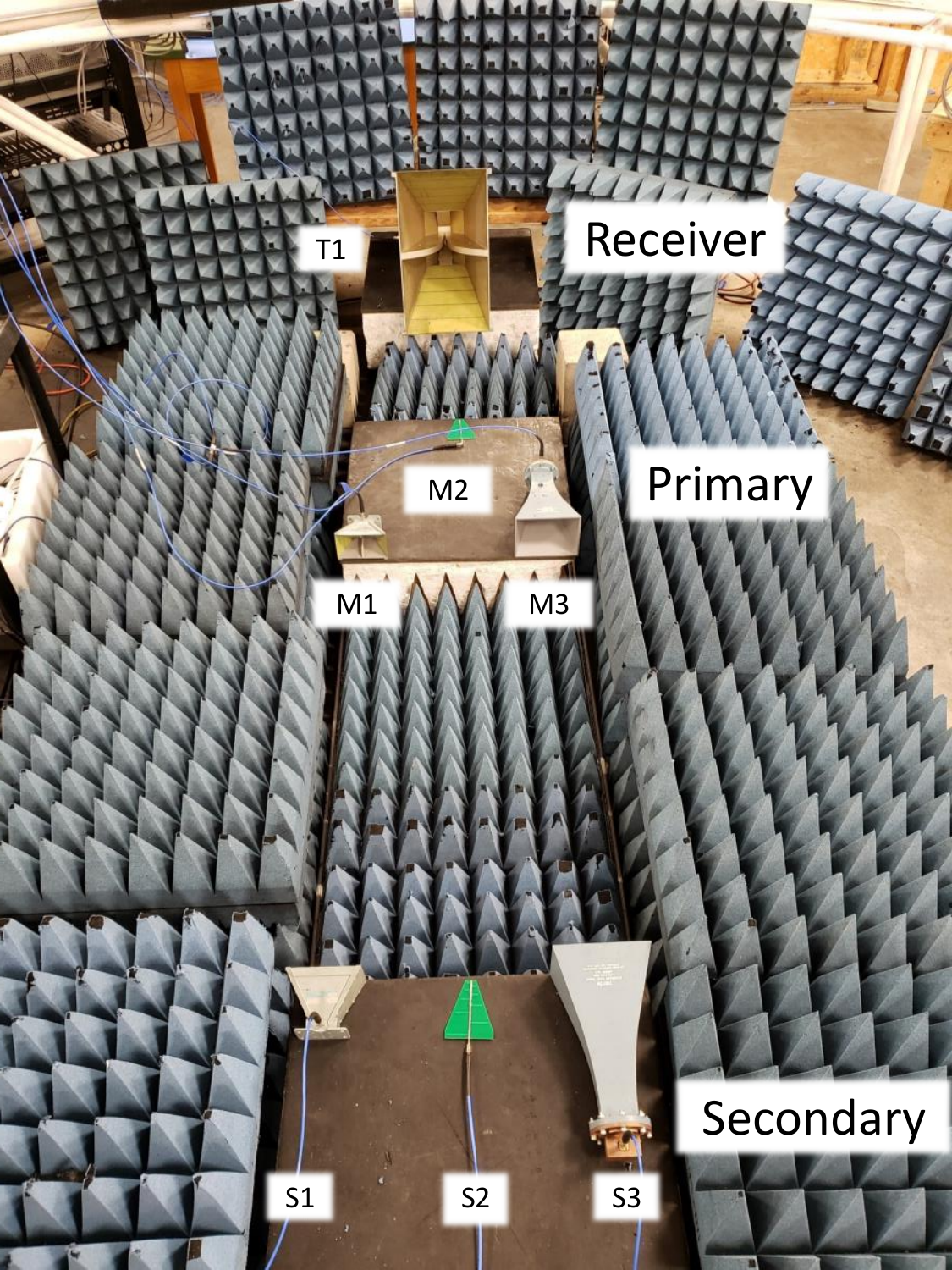}
	\textbf{c}
	\includegraphics[width=0.5\textwidth]{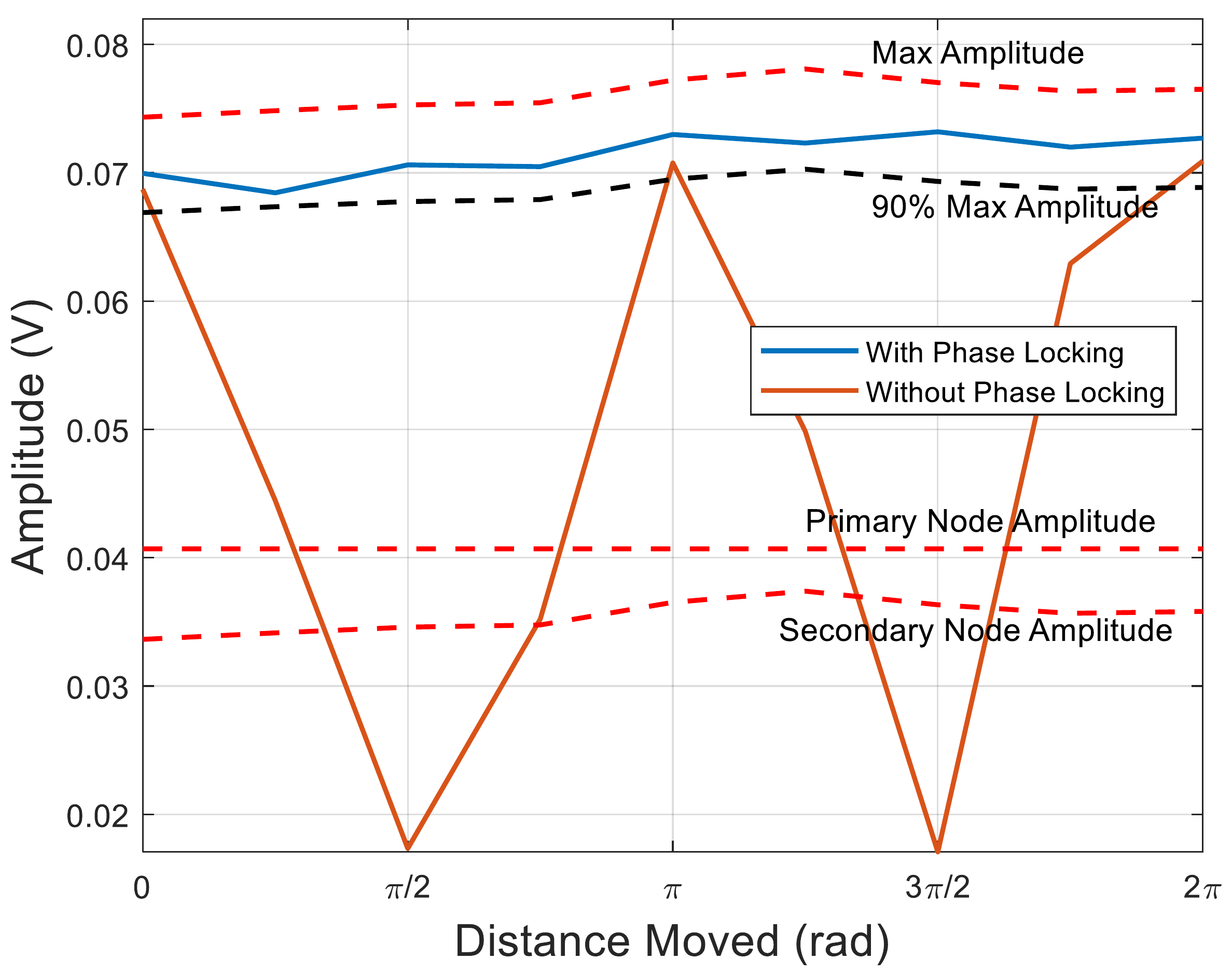}
	\caption[Optional caption]{\textbf{Experimental results of the two-node beamforming experiment with node motion.} \textbf{a.} Schematic of the two-node open-loop distributed beamforming experiment. \textbf{b.} Set in a semi-enclosed antenna range, this image shows the receiver, and the primary and secondary nodes. \textbf{c.} The dashed red lines at the bottom show the amplitudes of the two transmitted signals individually. The top dashed line shows the ideal signal gain assuming perfect summation of the two individual signals. The orange line shows the result of synchronizing the frequency but not the phase; two nulls appear, aligning to the total expected phase shift over the length of the motion of $720^\circ$. Outside this range, the signal returns to a high value, indicating successful frequency synchronization. The blue line shows both frequency and phase synchronization, demonstrating open-loop beamforming while the node is moved. The resulting signal is above 90\% of the ideal gain throughout.}
	\label{fig:results_wireless}
\end{figure}

%The coherent distributed beamforming setup described in this paper was validated through a wireless experiment. 
The block diagram along with an image of the experimental setup in a semi-enclosed arch range can be seen in Fig. \ref{fig:results_wireless}.  %For the wired experiment, antennas T1, M2, and S2 were replaced by cables, where the signals coming from the secondary and primary nodes were directly fed to an oscilloscope and then their signals were combined. The experimental setup is shown in Fig. \ref{fig:expSetup}.
 The synchronization signals were transmitted using standard gain horn antennas where M1 and S1 had an operational frequency range of 2-18~GHz, while the antennas used for M3 and S2 had an operational frequency range of 3.95-5.85~GHz. Two log-periodic antennas with an operational frequency range of 1.35-9.5~GHz were used to for the coherent beamforming of the nodes. The target receiver consisted of an oscilloscope connected to a horn antenna with operational frequency range of 0.5-6 GHz. The primary node was equipped with an active repeater that captures the ranging waveforms from the secondary node, amplifies them, and then retransmits them back to the secondary node. By doing this the propagation losses are proportional to $1/R^2$ rather than $1/R^4$ as seen in traditional radar. This will ensure that the desired signal will dominate any multipath. The center frequencies of the transmit and receive channels of the repeater are separated so that they lie far outside the instantaneous bandwidth that the radios can achieve. This serves to guarantee that any crosstalk between transmit and receive will be eliminated. The ranging waveform is transmitted from the secondary node using S1 at a center frequency of 3~GHz and retransmitted from M3 to S3 at a center frequency of 5~GHz. The frequency synchronization signals were also transmitted from the same horn as the ranging waveform where $f_{r1}$ and $f_{r2}$ were chosen as 4.3~GHz and 4.31~GHz respectively.

%The achievable coherent frequency was determined using \eqref{eq:highestFc}.
As described earlier, based on the SNR of the received ranging waveform, the positional uncertainty is expected to be 3.4~mm. Based on the analysis earlier can be shown that an uncertainty of less than $6$~mm is required for beamforming operations at 1.5 GHz, which is supported by the described ranging parameters. The inter-node distance is determined by an estimation of time of flight of the received signal on the secondary node found by a matched filtering process. The peak of the matched filter was interpolated with one thousand points using a built-in LabVIEW spline interpolation function to improve the accuracy. 
%Larger interpolation values proved to be too computationally expensive for the small amount of added performance. % The calculated ranging standard deviation was around 5 mm, then $f_c$ was chosen using \eqref{eq:highestFc}; to achieve a coherent action at $f_c = 1.5$~GHz the ranging standard deviation should be at most 6~mm, thus choosing this frequency was a suitable option. The coherent action was performed at $f_c = 1.5$~GHz which has a wavelength $\lambda_c = 0.2$~m for both the wired and wireless cases.
 The steering angle $\theta$ was set to $90^\circ$ since this represents the most challenging case for a two node system as shown in Fig.~\ref{fig:PGC}.

At the beginning of the test, the static phase offset $\phi_0$ was calibrated; this calibration represents a one-time calibration performed when the system is powered on and is generally not required later. The data was recorded over a total distance of $\lambda_c$ (20 cm), which was enough to evaluate the behavior of the distributed system, since the response will be repetitive every $\lambda_c$. Data was collected for 1,500 cycles of the beamforming frequency in 2~cm increments as the secondary node was moved way from the primary node. The collected data included the amplitude of the signals transmitted from S2 and M2, the amplitude of the signal summation for the cases where no phase correction was done, and where $\Delta \phi_{c}$ was used for phase correction. Two metrics were used to analyze the captured data: maximum achievable amplitude, and 90\% of the maximum achievable amplitude. 

The results of the wireless beamforming experiment are shown in Fig.~\ref{fig:results_wireless}. 
%The entire secondary node was moved a total of $\lambda_c$ in the opposite direction of the receiver. 
%\eqref{eq:Dpc} was used to correct the induced phase shifts.
 Since the beam steering angle $\theta$ was set to $90^\circ$, the total expected phase shift over the length of the motion is equal to $720^\circ$. This can be seen in the uncorrected data without phase locking, where the received signal undergoes two nulls over the length of the node motion. The received signal returns to the peak value between and outside of the nulls, indicating that the wireless frequency synchronization approach successfully maintained a frequency lock between the nodes. The blue line shows the case where the phase was updated continuously using the ranging estimate to adjust for node motion. The results demonstrate the ability to maintain greater than 90\% of the ideal beamforming gain consistently throughout the measurement.
\textbf{}

\textbf{Conclusions}

We have reported on the first fully wireless open-loop phase-coherent distributed beamforming experiment with relative node motion. The distributed system was demonstrated using a primary/secondary architecture to synchronize the frequencies of the nodes and to adjust the beamforming phase in response to node motion. 
%This was demonstrated through a wireless beamforming experiment at 1.5~GHz in a dynamic array where one of the nodes was displaced relative to the other with a beam steering angle $\theta = 90^\circ$. 
Frequency synchronization was achieved using an adjunct self-mixing analog circuit that generated an output reference frequency that was fed to a PLL internal to USRP X310 SDRs. Phase alignment was achieved using a high accuracy estimate of inter-node separation given by a TTSFW ranging approach. This method proved to provide more than 90\% coherent gain at each measured value of inter-node separation. 
Distributed beamforming was implemented in an open-loop format, without active feedback from the destination maintaining frequency and phase synchronization.
By demonstrating the ability to maintain open-loop distributed beamforming wirelessly between moving nodes, the results support the feasibility of implementing fully distributed wireless operations, including remote sensing, radar, and communications.
%. Two-tone signals were used to achieve frequency synchronization and high accuracy range measurements. The presented system is scalable once the two-tone stepped frequency waveform is used.

%****************************************************************************
\textbf{}

\textbf{Methods}

The experiment demonstrating open-loop phase-coherent distributed wireless system with dynamic nodes was held in a semi-enclosed arch range. Pyramidal foam absorbers were placed around the transmitters and receivers to reduce the effect of multipath RF reflections. 
%Multipath reflections, especially in indoor environments, can hinder the accuracy of phase estimation; slight phase shifts in the received frequency synchronization signals are amplified by the RF multipliers in the SDRs while generating the desired carrier frequency, thus a small phase shift for the reference signal translates to a large shift at the transmitted frequency. 
%A primary and a secondary node were used, the secondary node was synchronizing its frequency and phase to that of the primary node. 
Frequency and phase synchronization were performed using two antennas per node: one horn antenna with operational frequency range of 2-18 GHz denoted either by M1 or S1, and another horn antenna with operational frequency range of 3.95-5.85 GHz denoted either by M3 or S3. In addition to the synchronization antennas, each node was transmitting the coherent signals using a log periodic antenna with operational frequency range of 1.35 GHz-9.5 GHz denoted either by M2 or S2. The target consisted of an MSOX92004A Infiniium oscilloscope connected to a horn antenna with operational frequency range of 0.5-6 GHz which was denoted by T1.
The primary node was built using three USRP X310 SDRs; one SDRs was used to transmit the frequency synchronization signals, another one was used to transmit the coherent signals, and the last SDR was used as a repeater that amplifies and retransmits the received ranging signals. The secondary node was built using two USRP X310 SDRs and a frequency locking circuit; one SDR was used to transmit the coherent signals while the second one was used for ranging. All the SDRs were equipped with two UBX 160 USRP daughterboards. Every daughterboard was able to transmit and receive signals simultaneously from 10 MHz up to 6 GHz with an instantaneous bandwidth of 160 MHz. The antennas were connected using 2 m coaxial cables that support RF signals up to 18 GHz.
The frequency locking circuit in Fig. \ref{fig:waveform} was built using the following Mini-Circuits RF components: one ZVBP-4300+ cavity band pass filter with a bandwidth of 4.25 to 4.35 GHz, one VBFZ-5500-S+ band pass filter with a bandwidth of 4.9 to 6.2 GHz, one BLP-10.7+ low pass filter with a cutoff frequency of 11 MHz, two ZX60-83LN12+ LNAs with a gain of 22~dB and a bandwidth of 0.5--8~GHz, two ZFL-500HLN+ LNAS with a gain of 20 dB and a bandwidth of 10--500~MHz, one 10 dB attenuator, two ZX10-2-852-S+ power splitters with a bandwidth of 0.5--8.5~GHz, one ZX90-2-36-S+ multiplier with a bandwidth of 3.4--3.7~GHz. The TDK-Lambda LS25-12 was used to supply a 12 V power to the LNAs. The input to the oscilloscope was filtered using the VBFZ-1690-S+ band pass filter with a bandwidth of 1455--1925~MHz. The output of M1 was connected to the VLP-41 low pass filter with a cutoff frequency of 3.3 GHz. 
%The use of filters at the input ports of the SDRs was not necessary but it was done as an extra measure.
The primary node was equipped with an active repeater that captures the ranging waveforms from the secondary node, amplifies them, and then retransmits them back to the secondary node. In this manner, the propagation losses are proportional to $1/R^2$ rather than $1/R^4$ as seen in traditional radar, ensuring that the desired signal will dominate any multipath. The center frequencies of the transmit and receive of the repeater are separated so that they lie far outside the instantaneous bandwidth that the radios can achieve to guarantee that any crosstalk between transmit and receive will be eliminated. The ranging waveform is transmitted from the secondary node using S1 to M1 at a center frequency of 3~GHz and retransmitted from M3 to S3 at a center frequency of 5~GHz. The frequency synchronization signals were also transmitted from M3 where $f_{r1}$ and $f_{r2}$ were chosen as 4.3~GHz and 4.31~GHz respectively.
The positional uncertainty of the system is 3.4~mm, below the $6$~mm required for beamforming operations at 1.5 GHz, which is supported by the described ranging parameters. The inter-node distance is determined by an estimation of time of flight of the received signal on the secondary node found by a matched filtering process. The peak of the matched filter was interpolated with one thousand points using a built-in LabVIEW spline interpolation function to improve the accuracy. 
%All processing was performed in real time using LabVIEW 2018. 
The steering angle $\theta$ was set to $90^\circ$ since this represents the most challenging case for a two node system as shown in Fig.~\ref{fig:PGC}. 
%Selecting a fixed steering angle alleviates the need to perform angle estimation to localize the secondary nodes. 
At the beginning of the test, the static phase offset $\phi_0$ was calibrated; this calibration represents a one-time calibration performed when the system is powered on and is generally not required later. The data was recorded over a total distance of $\lambda_c$ (20 cm away from the primary node), which was enough to evaluate the behavior of the distributed system, since the response will be repetitive every $\lambda_c$. Data was collected for 1,500 cycles of the beamforming frequency in 2~cm increments as the secondary node was moved way from the primary node. The collected data included the amplitude of the signals transmitted from S2 and M2, the amplitude of the signal summation for the cases where no phase correction was done, and where $\Delta \phi_{c}$ was used for phase correction. The amplitudes were recorded by first transmitting the signal only from the primary node, then from the secondary node, afterwards both nodes were transmitting simultaneously with and without phase correction. Two metrics were used to analyze the captured data: maximum achievable amplitude, and 90\% of the maximum achievable amplitude.

\bibliographystyle{IEEEtran}

\end{document}